\documentclass[prl,aps,twocolumn,superscriptaddress,preprintnumbers,amsmath,amssymb]{revtex4}

\begin{document}
\preprint{DFPD-2014/TH/02}
\title{Geometrical hierarchies in classical supergravity}
\author{Hui Luo}
\affiliation{Dipartimento di Fisica ed Astronomia, Universit\`a di Padova, Via Marzolo 8, 35131 Padova, Italy}
\affiliation{INFN, Sezione di Padova, Via Marzolo 8, 35131 Padova, Italy}
\affiliation{SISSA, Via Bonomea 265, 34136 Trieste, Italy}
\author{Fabio Zwirner}
\affiliation{Dipartimento di Fisica ed Astronomia, Universit\`a di Padova, Via Marzolo 8, 35131 Padova, Italy}
\affiliation{INFN, Sezione di Padova, Via Marzolo 8, 35131 Padova, Italy}
%
\date{March 19, 2014}
\begin{abstract}
We introduce a $N=1$ supergravity model with a very simple hidden sector coupled to the electroweak gauge and Higgs sectors of the MSSM.  At the classical level, supersymmetry and $SU(2) \times U(1)$ are both spontaneously broken, with vanishing vacuum energy. Two real flat directions control the two symmetry-breaking scales $m_{3/2}$ and $m_Z$. The two massless scalars are a gauge singlet and the standard Higgs boson. All other unobserved particles have masses of order $m_{3/2}$. This may be a new starting point for studying the compatibility of naturalness with the observed mass hierarchies.  
\end{abstract}
%
\maketitle
%
%
\section{Introduction}
The 7-8 TeV run of the LHC \cite{moriond} saw the historical discovery of a scalar particle, with mass close to $125 \, {\rm GeV}$ and so far compatible with the Standard Model (SM) Higgs boson. 
Its highlights also include considerably stronger bounds on supersymmetric particles and additional Higgs bosons than those previously established at the LEP and Tevatron colliders. 
These results, complemented by other precise results from flavor physics, are challenging the concept of naturalness and its application to the gauge hierarchy problem.
At face value, the simplest supersymmetric extensions of the SM, for example the Minimal Supersymmetric Standard Model (MSSM), look fine-tuned with more than per-cent precision. 
However, there are good reasons to insist on the hypothesis  that supersymmetry plays a r\^ole in some unified theory of the fundamental interactions underlying the SM. 

Waiting for the sharper experimental picture that should emerge from the 13-14 TeV run of the LHC, theorists are broadening the spectrum of supersymmetric models under consideration. 
Two main approaches are being pursued \cite{av}: the first insists on the concept of naturalness and on a light spectrum of supersymmetric particles at the verge of being excluded; the second gives up the concept of naturalness and splits the mass scale of supersymmetric particles and extra Higgs bosons from the mass scale of the observed weak and Higgs bosons. 

It would be very important to establish whether this bifurcation is really unavoidable, especially if no new particle is discovered after LHC-14: is there some special supersymmetric extension of the SM that can solve the naturalness problem despite a little (or not-so-little) hierarchy between the weak scale and the scale of the so far unobserved MSSM particles? 
At present we do not have convincing examples, and it seems unlikely that a positive answer can be found sticking to renormalisable theories with softly broken rigid supersymmetry.  

Some inspiration may come from supergravity, where in addition to the gauge hierarchy problem we must also address the vacuum energy problem, and both problems are visible already at the classical level, when the spontaneous breaking of supersymmetry and of the electroweak gauge symmetry are implemented. 

In this Letter we perform a first step in the search for a $N=1$ supergravity model that, once embedded in a suitable ultraviolet completion, might be naturally compatible with the mass hierarchies imposed on us by experimental data. 
The model contains a very simple hidden sector, a chiral multiplet and a vector multiplet, where spontaneous supersymmetry breaking takes place according to the mechanism recently formulated in \cite{dz}. 
We include in the observable sector only the electroweak gauge sector and the Higgs sector of the MSSM, leaving aside for the moment, for the sake of simplicity, the matter sector and the strong interactions.  

At the classical level, and because of its geometrical properties, the model exhibits some remarkable features: 
supersymmetry and the gauge symmetry are both spontaneously broken, with vanishing vacuum energy; 
two independent real flat directions control the scale of supersymmetry breaking in Minkowski space, parametrised by the gravitino mass $m_{3/2}$, and the scale of electroweak gauge symmetry breaking, parametrised by the weak boson masses $m_{W,Z}$; 
apart from the massive weak bosons, the photon and two classically massless scalars, a gauge singlet and a SM-like Higgs boson, all the other states in the spectrum do/can \cite{massfoot} have masses of order $m_{3/2}$; 
all renormalisable interactions are exactly as in the MSSM, with a definite prediction for several of its parameters, and the non-renormalisable interactions are suppressed by inverse powers of the Planck mass $M_P = (8 \pi G_N)^{-1/2} \simeq 2.4 \times 10^{18} \, {\rm GeV}$.

These results are a promising starting point, with some novel ingredients, for addressing the dynamical generation of the observed hierarchies of scales in supergravity models. 
To make the model realistic, however, we must include the strong interactions, the matter sector and the quantum corrections. 
Such programme deserves a separate study and goes beyond the aim of the present Letter: we conclude by outlining some of the open questions and a possible future strategy for addressing them.

\section{The model}
We now specify our model, which couples the hidden sector of \cite{dz} to the electroweak gauge and Higgs sector of the MSSM, in the standard formalism of $N=1$, $d=4$ supergravity \cite{wb} and in natural units where $M_P=1$. 

The gauge group is $SU(2)_L \times U(1)_Y \times \widetilde{U(1)}$. 
The first two factors are associated with electroweak interactions, as in the SM.
The vector multiplet ${\cal V} \sim (\widetilde{V}, V_\mu)$ of $\widetilde{U(1)}$ is part of the hidden sector that breaks supersymmetry. 

The chiral multiplets are a SM-singlet, ${\cal T} \sim (T,\widetilde{T})$, and the two MSSM Higgs doublets, ${\cal H}_1 \sim (H_1, \widetilde{H}_1)$ and  ${\cal H}_2 \sim (H_2, \widetilde{H}_2)$. 
A crucial feature, inherited from \cite{dz}, is that the imaginary part of $T$ shifts under $\widetilde{U(1)}$, whilst the two Higgs superfields do not transform: 
\begin{equation}
\delta_\epsilon T = i \, \epsilon \, , 
\qquad
\delta_\epsilon H_1 = \delta_\epsilon H_2 = 0 \, ,
\qquad
(\epsilon \in \mathbf{R}) \, .
\end{equation} 

Motivated by string compactifications and by extended supergravities \cite{fgkp}, we choose the K\"ahler manifold for the scalar fields (unifying the chiral multiplets in the hidden and Higgs sectors) to be $SO(2,5)/[SO(2) \times SO(5)]$:
\begin{equation}
e^{-K} = ( T + \overline{T})^2 - | H_1^0 - \overline {H_2^0} |^2 - |H_1^- + \overline{H_2^+} |^2 \, ,
\label{kahler}
\end{equation}
and, in the field basis of (\ref{kahler}), a constant superpotential:
\begin{equation}
W = \sqrt{2} \, \widetilde{g} \, .
\label{superp}
\end{equation}

Finally, we choose a factorised gauge kinetic function:
\begin{equation}
\widetilde{f} = \frac{1}{ \widetilde{g}^{\, 2}} \, , 
\quad
f_Y = a_Y + b_Y \, T \, , 
\quad
f_L = a_L + b_L \, T \, ,
\label{gkinf}
\end{equation}
where $(\widetilde{g},a_Y,b_Y,a_L,b_L)$ are real constants. As explained in \cite{dz}, the fact that $W$ and $\widetilde{f}$ are controlled by the same coupling constant $\widetilde{g}$ is not a fine-tuning, but the consequence of an underlying $N=2$ gauged supergravity.

\section{Classical potential and vacua}
The classical potential of the model is:
\begin{eqnarray}
V_0 & = &  e^{2 K} \, (A+B+C+D) \, , 
\label{v0}
\\
A & = & 2 \, \widetilde{g}^2 \, ( | H_1^0 - \overline {H_2^0} |^2 + |H_1^- + \overline{H_2^+} |^2 )
\nonumber \\
& = & 2 \, \widetilde{g}^2 \, [ H_1^\dagger H_1 + H_2^\dagger H_2 - ( H_1 H_2 + {\rm h.c.} ) ]
\label{potA}
\, , \\
B & = & \frac{g^{\, \prime \, 2}}{8} \, (  | H_1^0 |^2 - | H_2^0 |^2 + |H_1^- |^2  - | H_2^+ |^2 )^2
\, , \\
C & = & \frac{g^{2}}{2}  \, | H_1^0  \, \overline {H_1^-} + \overline{H_2^0} \, H_2^+ |^2 
\, , \\
D & = & \frac{g^2}{8}  \, (  | H_1^0 |^2 - | H_2^0 |^2 - |H_1^- |^2  + | H_2^+ |^2 )^2 \, ,
\label{dd}
\end{eqnarray}
where we have introduced the field-dependent $SU(2)_L$ and $U(1)_Y$ coupling constants:
\begin{equation}
g^{\, \prime \, 2} \equiv \frac{1}{Re \, f_Y} \, , 
\qquad
g^2 \equiv \frac{1}{Re \, f_L} \, . 
\end{equation}

Each of the four addenda contributing to $V_0$ is positive semidefinite. After gauge fixing, inequivalent vacua can be classified by $\langle H_1^- \rangle = \langle H_2^+ \rangle = 0$ and
\begin{equation}
\label{vevs}
\langle T \rangle = x \, , 
\qquad
\langle H_1^0 \rangle = \langle H_2^0 \rangle = 2 \, x \, v \, ,
\end{equation}
where $x>0$ and $v \ge 0$ parametrise two real flat directions. As in \cite{dz}, the $\widetilde{U(1)}$ gauge symmetry and supersymmetry are spontaneously broken on flat Minkowski space at all vacua. The electroweak gauge symmetry is also spontaneously broken on the generic vacuum, although it can be restored at the special point $v=0$. 

\section{Spectrum and interactions}

In the hidden sector, the spectrum is exactly as in \cite{dz}:
\begin{equation}
m_{3/2}^2 = m_{1/2}^2 = \frac{\widetilde{g}^2}{2 \, x^2} \, ,
\quad
m_V^2 = 2 \, m_{3/2}^2 \, ,
\quad
m_0^2 = 0 \, ,
\label{hidspec}
\end{equation}
in a self-explanatory notation. Setting 
\begin{equation}
T = x \, (1 + t + i \, \tau) \, , 
\label{tdef}
\end{equation}
$\tau$ is the Goldstone boson absorbed by the massive vector $V_\mu$, and $t$ is a canonically normalised massless scalar. Similarly, the Goldstino absorbed by the massive gravitino is a linear combination of $\widetilde{T}$ and $\widetilde{V}$, and the orthogonal combination is a massive spin-1/2 Majorana fermion. 

In the observable sector, the spectrum corresponds to a special choice of parameters in the MSSM. The gauge boson masses are: 
\begin{equation}
m_\gamma^2 = 0 \, , 
\quad
m_W^2 = \overline{g}^2 \, v^2 \, , 
\quad
m_Z^2 =  (\overline{g}^2 + \overline{g}^{\, \prime \, 2})  \, v^2 \, ,
\end{equation}
where $\overline{g} \equiv \langle g \rangle$ and $\overline{g}^{\, \prime} \equiv \langle g^{\, \prime} \rangle$. The Higgs boson spectrum can be easily obtained from $V_0$ by performing the following decomposition, which brings all the kinetic terms to canonical form and diagonalises all the mass terms:
\begin{eqnarray}
\label{prima}
H_1^- & = & \sqrt{2} \, x \,  (H^- - G^-) \, ,
\\
H_2^+ & = & \sqrt{2} \, x \, (H^+ + G^+) \, ,
\\
H_1^0 & = & 2 \, x \left( v + \frac{h^0+H^0}{2} + i \, \frac{A^0 - G^0}{2} \right) \, ,
\\
H_2^0 & = & 2 \, x \left( v + \frac{h^0-H^0}{2} + i \, \frac{A^0 + G^0}{2} \right) \, .
\label{ultima}
\end{eqnarray}
The result is:
\begin{equation}
m_A^2 = 2 \, m_{3/2}^2 \, , 
\quad
m_{\pm}^2 = m_A^2 + m_W^2 \, ,
\label{ahpm}
\end{equation}
\begin{equation}
m_h^2 = 0 \, , 
\qquad
m_H^2 = m_A^2 + m_Z^2 \, .
\label{shbh}
\end{equation}
In MSSM notation, see (\ref{potA}), it corresponds to 
\begin{equation}
\label{potpar}
m_1^2=m_2^2=-m_3^2= m_{3/2}^2 \, , 
\quad
\left( \beta = - \alpha = \frac{\pi}{4} \right) \, . 
\end{equation}
Notice that here the relations (\ref{potpar}) follow from the classical K\"ahler geometry (\ref{kahler}) and not from a fine-tuning.

In the gaugino/higgsino sector, and in a suitable basis of canonically normalised fields, the chargino and neutralino mass matrices are as in the MSSM, with $\beta=\pi/4$:  
\begin{equation}
{\cal M}_C = \left(
\begin{array}{cc}
M_2 &  m_W  \\
m_W  & \mu
\end{array}
\right) \, ,
\end{equation}
\begin{equation}
{\cal M}_N = \left(
\begin{array}{cccc}
M_1 & 0 & - \frac{m_Z s_W}{\sqrt{2}} & \frac{m_Z s_W}{\sqrt{2}}  \\
0 & M_2 &   \frac{m_Z c_W}{\sqrt{2}}  & - \frac{m_Z c_W}{\sqrt{2}}  \\
- \frac{m_Z s_W}{\sqrt{2}}  & \frac{m_Z c_W}{\sqrt{2}}  & 0 & - \mu  \\
\frac{m_Z s_W}{\sqrt{2}}  & - \frac{m_Z c_W}{\sqrt{2}}  & - \mu & 0  \\
\end{array}
\right) \, .
\end{equation}
In the above equations, $s_W \equiv \sin \theta_W$ and  $c_W \equiv \cos \theta_W$. The Higgsino mass parameter is
\begin{equation}
\label{muterm}
\mu=m_{3/2} \, .
\end{equation}
Notice that, in contrast with the MSSM, the superpotential (\ref{superp}) does not contain a  Higgs mass term. However, an effective $\mu$-term is generated from the K\"ahler potential, according to a well-known mechanism of broken supergravity \cite{gm}, first explored in \cite{so2mu} for the special K\"ahler manifold of (\ref{kahler}). The gaugino mass parameters are
\begin{equation}
\label{gaumass}
M_1 = m_{3/2} \, (1 - g^{\prime \, 2} \, a_Y) \, , 
\quad
M_2 = m_{3/2} \, (1 - g^2 \, a_L) \, . 
\end{equation}
Two extreme choices for the gauge kinetic functions $f_Y$ and $f_L$ in (\ref{gkinf}) are worth considering. The first one, corresponding to $b_Y = b_L=0$, leads to constant $f_Y=a_Y=1/g^{\, \prime \, 2}$ and  $f_L=a_L=1/g^2$, thus to $M_1=M_2=0$ \cite{gaunote}. The second one, $a_Y=a_L=0$, leads to $M_1=M_2=m_{3/2}$. 

After moving to canonically normalised fields and taking the appropriate flat limit, the model factorises into a decoupled hidden sector times the electroweak gauge and Higgs sectors of the MSSM: all the MSSM renormalisable interactions are reproduced \cite{flatnote}, for  the parameter choices (\ref{potpar}), (\ref{muterm}) and (\ref{gaumass}).

Keeping $M_P$ finite, the low-energy effective Lagrangian includes, besides the MSSM, the gravitational interaction and other supergravity interactions, corresponding to local operators of dimension $d \! > \! 4$, suppressed by $M_P^{4-d}$.  

\section{Discussion}

Our model realises, in an economical and predictive framework, some features previously discussed in supersymmetric extensions of the SM, but never combined. 

The breaking of supersymmetry with vanishing classical vacuum energy and the gravitino mass sliding along a classical flat direction is the feature of no-scale models \cite{cfkn}. 
The additional breaking of the electroweak gauge symmetry, along another classical flat direction and preserving the vanishing of the classical vacuum energy, was previously introduced in \cite{bz}.  
There, however, additional classical flat directions were present, both in the hidden sector and in the MSSM Higgs sector. 
Here, instead, the only two classical flat directions are in one-to-one correspondence with the scales of supersymmetry and electroweak symmetry breaking. 
The axion $\tau$ is absorbed by the massive $\widetilde{U(1)}$ vector, as in \cite{dz}. 
The only classically massless fields are the dilaton $t$ in the hidden sector and the SM Higgs boson $h$ \cite{hnote} in the MSSM Higgs sector.
The masses of the other Higgs bosons receive supersymmetry-breaking contributions of order $m_{3/2}$. 

It would be natural to interpret the two massless scalars as pseudo-Goldstone bosons of some accidental symmetry.
Such an interpretation is possible, with some qualifications. 
All the isometries of the K\"ahler manifold (\ref{kahler}) other than the gauged ones are explicitly broken by the potential (\ref{v0}). 
However, after moving to the unitary gauge $\tau=0$ for the gauged shift symmetry, $H_1^0 - \overline{H_2^0}=H_1^- + \overline{H_2^+} =0$ (i.e. $A^0=H^0=H^\pm=0$) solve their classical equations of motion for arbitrary configurations of the remaining fields. 
The truncated scalar Lagrangian, obtained by inserting the above solutions in the original one, is expressed in terms of the residual scalars $(T,H_1^0 +\overline{H_2^0},H_1^--\overline{H_2^+})$ and has two global symmetries spontaneously broken on the vacuum.
The masslessness of $t$ is accounted for by the rigid scale transformations:
\begin{equation*}
(T,H_1^0 + \overline{H_2^0},H_1^--\overline{H_2^+}  ) \rightarrow \rho \, (T,H_1^0 + \overline{H_2^0},H_1^--\overline{H_2^+}  )  \, ,
\end{equation*}
($\rho \in \mathbf{R}$).
As already discussed in \cite{hkw}, the masslessness of $h$ is accounted for by the rigid shift symmetry ($\sigma \in \mathbf{R}$):
\begin{equation*}
H_1^0 + \overline{H_2^0} \rightarrow H_1^0 + \overline{H_2^0} + \sigma \, .
\end{equation*}

\section{Variations}

We briefly describe some possible variations on the model discussed above, concerning the gauge group and the manifold for the scalar fields. 
They weaken the connection of the model with string compactifications and extended supergravity, in particular the geometrical explanation of (\ref{potpar}), but they preserve some other remarkable properties, with some differences that may play a r\^ole in the search for realistic completions. 

The manifold for the scalar fields changes from the one in (\ref{kahler}) to the one described by the K\"ahler potential
\begin{equation}
\widehat{K} = - 3 \, \log (T + \overline{T}) + \frac{| H_1^0 - \overline {H_2^0} |^2 + |H_1^- + \overline{H_2^+} |^2}{(T + \overline{T})^{n}} \, , 
\label{khat}
\end{equation}
with $n \le 1$ or $n \ge 2$ (not necessarily integer). 
Simultaneously, the gauge group becomes $SU(2)_L \times U(1)_Y$, without the $\widetilde{U(1)}$ vector multiplet gauging the $T$ shift symmetry. 
The superpotential $W$ and the gauge kinetic functions $f_Y$ and $f_L$  remain the same as in (\ref{superp}) and (\ref{gkinf}). 

The classical potential is still positive semi-definite for conceivable values of $n$ and of the ratio appearing in (\ref{khat}). It has now with a complex $T$ flat direction in addition to the real flat direction associated with the SM Higgs field. In the hidden sector, the spectrum consists of the massless dilaton $t$ and axion $\tau$ associated with the $T$ fluctuations, and of the massive gravitino, which absorbs the Goldstino $\widetilde{T}$, with $m_{3/2}^2 = \widetilde{g}^{\, 2}/(4 \, x^3)$ if we stick to (\ref{tdef}). In the MSSM sector, after replacing $2x$ with $(2x)^{n/2}$ on the right-hand side of (\ref{vevs}) and in (\ref{prima})--(\ref{ultima}), the spectrum is as before, with the two modifications:
\begin{equation}
m_A^2 = 2 \, (n-2) \, (n-1) \, m_{3/2}^2 \, , 
\quad
\mu=(1-n) \, m_{3/2} \, .
\end{equation} 

\section{Outlook}

We regard the results of this Letter as a novel and promising starting point for investigating the dynamical generation, {\em \`a la} Coleman-Weinberg \cite{cw}, of the observed hierarchies of scales and of a phenomenologically viable spectrum in realistic supergravity models. 
However, several steps need to be performed to carry out such an investigation. 

First, we should complete the MSSM gauge group by including also the $SU(3)_C$ factor associated with the strong interactions: this is straightforward, the only freedom being the choice of a gauge kinetic function $f_C$ similar in form to those in (\ref{gkinf}).  

Then, we should include the MSSM matter sector. 
This is also straightforward in principle, but choosing the K\"ahler potential for the matter fields introduces some arbitrariness in model building. 
The problem is simplified by the fact that, in realistic models, we can expand the K\"ahler potential up to quadratic fluctuations in the quark and lepton superfields. 
However, the $T$-dependence of the coefficients will affect the spectrum, in particular the supersymmetry-breaking mass terms. 
Moreover, the breaking of the global symmetries by the standard top Yukawa coupling in the superpotential might be too hard for a natural generation of the desired mass hierarchies and spectrum. 
It may be interesting to explore the possibility of generating the top quark mass by mixing with heavy fermions, in analogy with models of partial compositeness \cite{partcomp,susypc}. 

Finally, we can envisage computing the calculable (logarithmic) quantum corrections in the model, suitably parametrising those that would require the knowledge of the ultraviolet completion of our effective supergravity, with the goal of checking whether realistic mass hierarchies can be generated for some values of the parameters.  

The first attempts at carrying out this program were performed in \cite{qnoscale}, including only the quantum corrections associated with the MSSM fields, dismissing the possibility of ${\cal O} (m_{3/2}^2 M_P^2)$ contributions to the effective potential, signalled by the one-loop quadratically divergent contributions proportional to ${\rm Str} \, {\cal M}^2$, and ignoring also the ${\cal O} (m_{3/2}^4)$ cosmological term in the MSSM potential. 
Some of these issues were addressed later.
It was shown in \cite{fkz} that there are special classes of supergravity models, whose geometrical structure is inherited from superstring compactifications or from gauged extended supergravities, where the only field-dependence of  ${\rm Str} \, {\cal M}^2$ along the classically flat directions is via the gravitino mass, ${\rm Str} \, {\cal M}^2 = k \, m_{3/2}^2$, where $k \in \mathbf{R}$ is a constant. 
This may allow for cancellations of the one-loop quadratic divergences after the inclusion of all hidden sectors, and there are examples of models where $k=0$. 
It is encouraging that in the model considered here ${\rm Str} \, {\cal M}^2 = 0$ and  ${\rm Str} \, {\cal M}^2 = - 8 \, m_{3/2}^2$, respectively, for the two choices of gauge kinetic functions discussed after  (\ref{gaumass}).
The corresponding values for the variations are ${\rm Str} \, {\cal M}^2 =  - 4 \, (2n \mp 1) \, m_{3/2}^2$.
Assuming the absence of ${\cal O} (m_{3/2}^2 M_P^2)$ contributions, the possibility of generating the desired hierarchies was further discussed in some special cases.
The importance of the cosmological term in the MSSM potential was stressed in \cite{kpz}, where the implications of moduli-dependent Yukawa couplings for the third generation were also studied.
The possibility of explaining a little hierarchy between $m_{h,W,Z}$ and $m_{3/2}$ was pointed out in \cite{bs}. 

Despite these partial results, we feel that a novel systematic study of the radiative generation of the desired hierarchies, in the possible realistic completions of our model, is in order. 
It should take into account today's experimental constraints and the different options for giving a mass to the top quark and its superpartners. 
We are looking forward to addressing these questions in a future work. 

\section{Acknowledgments}

We thank G.~Dall'Agata, F.~Feruglio, G.~Villadoro and A.~Wulzer for useful discussions. H.L. and this work are supported by the ERC Advanced Grant no.267985 (\textit{DaMeSyFla}). 
%
%

%

\begin{thebibliography}{99}
\bibitem{massfoot}
More precisely, the masses of the extra Higgs bosons and the pure Higgsino mass terms are of order $m_{3/2}$, while pure gaugino mass terms can range from zero to order $m_{3/2}$, depending on the choice of gauge kinetic functions.
\bibitem{gaunote}
For large enough $m_{3/2}$, realistic gaugino masses could be generated radiatively.
\bibitem{flatnote}
However, it should be kept in mind that all MSSM mass parameters (and possibly the gauge couplings) depend on the flat direction $x$.
\bibitem{hnote}
A classically massless $h$ is not a (qualitative) problem: if quantum corrections eventually select $v$, a mass for $h$ is automatically generated.
\end{thebibliography}
\end{document}